\documentclass{article}
\usepackage{ijcai25}

\usepackage{times}
\usepackage{soul}
\usepackage{url}
\usepackage[hidelinks]{hyperref}
\usepackage[utf8]{inputenc}
\usepackage[small]{caption}
\usepackage{graphicx}
\usepackage{amsmath}
\usepackage{amsthm}
\usepackage{amssymb}
\usepackage{booktabs}
\usepackage{algorithm}
\usepackage{algorithmic}
\usepackage[switch]{lineno}

\title{Interpreting CFD Surrogates through Sparse Autoencoders}

\author{
Yeping Hu
\and
Shusen Liu\\
\affiliations
Lawrence Livermore National Laboratory\\
\emails
\{hu25, liu42\}@llnl.gov
}
\begin{document}

\maketitle

\begin{abstract}
Learning‑based surrogate models have become a practical alternative to high‑fidelity CFD solvers, but their latent representations remain opaque and hinder adoption in safety‑critical or regulation‑bound settings. This work introduces a post-hoc interpretability framework for graph-based surrogate models used in computational fluid dynamics (CFD) by leveraging sparse autoencoders (SAEs). By obtaining an overcomplete basis in the node embeddings space of a pretrained surrogate, the method extracts a dictionary of interpretable latent features. The approach enables the identification of monosemantic concepts aligned with physical phenomena such as vorticity or flow structures, offering a model-agnostic pathway to enhance explainability and trustworthiness in CFD applications.

\end{abstract}

\section{Introduction}

High‑fidelity computational fluid dynamics (CFD) simulations remain the gold standard for analyzing complex unsteady flows, but they are too expensive for real‑time control, design‑space exploration, or uncertainty quantification \cite{najm2009uncertainty}, \cite{hu2024comparative}. Graph‑based surrogate models \cite{pfaff2020learning}, \cite{hu2023graph} have recently emerged as promising alternatives, giving state‑of‑the‑art accuracy while achieving speedups of 11x-290x compared to traditional solvers \cite{beale1985high}. Unfortunately, these learning-based surrogate models are black boxes: their node‑level latent representations are high‑dimensional, entangled, and difficult to relate back to physical concepts such as coherent structures, pressure gradients, or vortex shedding. The absence of trustworthy explanations hampers industrial adoption and XAI compliance \cite{walke2023artificial}.

Sparse Autoencoders (SAEs) \cite{Cunningham2023}, \cite{Gao2024}, \cite{Marks2024}, \cite{Mudide2024}, \cite{Muhamed2024} have recently become a leading tool for mechanistic interpretability in large language models (LLMs). Wide, over‑complete, yet sparse autoencoder trained on hidden activations can discover thousands of potentially monosemantic features, each associated with a human‑readable concept.
% —while preserving downstream performance. 
To the best of our knowledge, SAEs have not been applied to post‑hoc interpretability of physics‑based surrogates. Likewise, recent efforts that seek to make surrogates interpretable rely on architectural changes or fine‑tuning with additional regularizers \cite{barwey2023interpretable} rather than post‑hoc dictionary learning.

This gap motivates our study. In this paper, we propose a post‑hoc pipeline that trains an SAE on the frozen node‑embeddings produced by a pretrained graph-based surrogate and uses the learned sparse dictionary to \textit{identify whether the learned embeddings can be explained and the corresponding surrogate is trustworthy}. Our contributions are listed below:
\begin{itemize}
    \item We introduce the first application of sparse autoencoders for post‑hoc, model‑agnostic interpretation of latent spaces in graph‑based CFD surrogate models.
    \item We conduct a comprehensive evaluation using physics‑driven criteria—including temporal consistency, spatial localization, and vortex-region alignment—showing that SAE yields interpretable and physically meaningful representations.
\end{itemize}

\section{Related Work}

\paragraph{Sparse Autoencoders} 
Sparse autoencoders (SAEs) have recently gained traction as tools to improve model interpretability. Cunningham et al.\ \cite{Cunningham2023} demonstrate that training SAEs on residual streams of language models yields highly interpretable, monosemantic features. Gao et al. \cite{Gao2024} scale SAEs to larger LLMs, finding clear scaling laws and introducing quantitative metrics to evaluate feature quality. Several improvements upon the original SAE setups have also been proposed, e.g., gated \cite{rajamanoharan2024improving} k-sparse \cite{makhzani2013k},  L0-regularization \cite{rajamanoharan2024jumping}, mutual feature regularization \cite{Marks2024}, and witch SAEs \cite{Mudide2024}. A curated set of pretrained SAEs is also available from Gemma-Scope \cite{lieberum2024gemma} on a wide variability of layers and model sizes.
In the vision domain, to improve cross-model interpretability, Thasarathan et al.\ \cite{Thasarathan2025} propose Universal SAEs that align concepts across different vision models.  Stevens et al. \cite{Stevens2025} use SAEs for causal interventions on vision features, enabling rigorous testing of model behavior. 
% The switch SAEs \cite{Mudide2024} improve computational scalability and enabling expert-based feature extraction with lower cost.
% Several works adapt SAEs to specialized interpretability tasks. 
% Muhamed et al.\ \cite{Muhamed2024} introduce SAEs for identifying rare ``dark matter'' concepts, while the PatchSAE work \cite{Lim2025} shows how vision model representations evolve with fine-tuning. 

% While Heap et al.\ \cite{Heap2025} caution that SAEs may discover interpretable features even in random models, the collective evidence supports their utility in XAI as scalable and concept-aligned tools for feature extraction and analysis.

\paragraph{Interpretation of CFD Surrogates}
On the application side, recent advancements in explainable artificial intelligence (XAI) have been pivotal in interpreting neural network models applied to scientific simulations involving spatial data. This section reviews related studies that have contributed to this domain.
% \textit{Saliency and Gradient-Based Explanations}
Amico et al.~\cite{amico2025data} applied gradient-based XAI methods, including Grad-CAM and SHAP, to a U-Net architecture predicting 2D velocity fields of turbulent jets. Their findings revealed that the network focused on moderate-vorticity regions, providing insights into turbulence dynamics.
% \textit{Feature Attribution Methods}
Mamalakis et al.~\cite{mamalakis2023carefully} discussed the importance of baseline selection in attribution methods like Integrated Gradients and SHAP when applied to climate data. They demonstrated that different baselines could lead to varying interpretations, emphasizing the need for careful baseline choice in geoscientific applications.
% \textit{Other XAI for Science Works}
Xiang et al.~\cite{xiang2025explainable} integrated hydrological principles into a recurrent neural network to simulate and forecast floods. 
% By embedding the model into the network architecture, they enhanced both the performance and interpretability of the model. 
In another study, Mamalakis et al.~\cite{mamalakis2023using} utilized XAI to assess the distinguishability of climate states before and after stratospheric aerosol injection. Their approach provided a data-driven framework to evaluate the impacts of climate interventions.

\section{Method}
\subsection{Graph-Based Surrogate for CFD Simulation}
We employ MeshGraphNets (MGN) \cite{pfaff2020learning} as the core surrogate model because they pair geometric flexibility on unstructured CFD meshes with state‑of‑the‑art predictive accuracy.  
MGN follows an encoder–process–decoder (EPD) paradigm.  
For a graph snapshot \(\mathcal{G}_t=(\mathcal{V},\mathcal{E})\) at time \(t\), each node \(i\in\mathcal{V}\) carries raw features \(\mathbf{x}_{i,t}\) (e.g.,\ velocity, pressure, coordinates), and each edge \((i,j)\in\mathcal{E}\) carries geometric attributes \(\mathbf{x}_{ij,t}\) (relative position, face normal, etc.).  
First, two multilayer perceptrons (MLPs) embed these quantities:

\begin{equation}
\mathbf{h}^{0}_{i,t}=f_n\!\bigl(\mathbf{x}_{i,t}\bigr), \qquad
\mathbf{h}^{0}_{ij,t}=f_e\!\bigl(\mathbf{x}_{ij,t}\bigr),
\end{equation}

\noindent
yielding initial node and edge representations \(\mathbf{h}^{0}_{i,t}\) and \(\mathbf{h}^{0}_{ij,t}\).  
Information is then propagated through \(L\) steps of message passing.  
At each step \(l\;(1\le l\le L)\) the embeddings are updated by

\begin{equation}
\mathbf{h}^{l}_{i,t}=f^{l}_{n}\!\Bigl(\mathbf{h}^{l-1}_{i,t}, \sum_{j\in\operatorname{Adj}(i)}\mathbf{h}^{l-1}_{ij,t}\Bigr), \
\mathbf{h}^{l}_{ij,t}=f^{l}_{e}\!\Bigl(\mathbf{h}^{l-1}_{ij,t},
\mathbf{h}^{l-1}_{i,t},\mathbf{h}^{l-1}_{j,t}\Bigr),
\end{equation}

\noindent
where \(\operatorname{Adj}(i)\) denotes the neighbors of node \(i\).  
After \(L\) iterations the node embeddings \(\mathbf{h}^{L}_{i,t}\) encode multi‑hop spatial interactions across \(\mathcal{G}_t\). The decoder is a third MLP, \(g_n\), that maps the processed node embeddings to the predicted next‑time‑step state (or to a state increment \(\Delta\mathbf{x}_{i,t}\)): $\widehat{\mathbf{x}}_{i,t+1}=g_n\!\bigl(\mathbf{h}^{L}_{i,t}\bigr)$. The network is trained end‑to‑end with a mean‑squared‑error objective:

\begin{equation}
\mathcal{L}_\text{MGN} = \frac{1}{|\mathcal{V}|}\sum_{i\in\mathcal{V}}
\bigl\|\widehat{\mathbf{x}}_{i,t+1}-\mathbf{x}_{i,t+1}\bigr\|_2^{2},
\end{equation}

\noindent
optimized over snapshots drawn from unsteady CFD simulations.  
%Once trained, we freeze the MGN parameters and treat the final node embeddings \(\mathbf{h}^{L}_{i,t}\) as the latent representations to be analyzed by our sparse autoencoder. %enabling post‑hoc interpretability of the surrogate without altering its predictive performance.

\subsection{Sparse‑Autoencoder Dictionary Learning}
To dissect the latent space produced by the frozen MGN, we train a sparse autoencoder (SAE) \cite{Gao2024} on the collection of node embeddings $\{\mathbf{h}^{L}_{i,t}\}$ from testing data.  Let \(d_{\text{in}}\) denote the dimensionality of a node embedding \(\mathbf{h}\in\mathbb{R}^{d_{\text{in}}}\) produced by the surrogate.  We choose an expansion factor \(\kappa>1\) and set the width of the hidden layer to \(d_{\text{hid}}=\kappa\,d_{\text{in}}\), yielding a set of over-complete bases that can isolate fine‑grained concepts. The encoder is a single linear map followed by a ReLU non‑linearity,
\begin{equation}
    \mathbf{z}
    =\sigma\!\bigl((\mathbf{h}-\mathbf{b}_{\mathrm{dec}})\,W_{\mathrm{enc}}+\mathbf{b}_{\mathrm{enc}}\bigr),
    \qquad
    W_{\mathrm{enc}}\in\mathbb{R}^{d_{\text{in}}\times d_{\text{hid}}},
\end{equation}
where \(\sigma(\cdot)=\max(0,\cdot)\). The decoder reconstructs the embedding via
\begin{equation}
    \widehat{\mathbf{h}}=\mathbf{z}\,W_{\mathrm{dec}}+\mathbf{b}_{\mathrm{dec}},
    \qquad
    W_{\mathrm{dec}}\in\mathbb{R}^{d_{\text{hid}}\times d_{\text{in}}}.
\end{equation}
Immediately after initialization and after every optimization step, the rows of \(W_{\mathrm{dec}}\) are re-normalised to unit \(\ell_{2}\) norm.  This constraint fixes the scale of each dictionary atom and empirically stabilizes training. 

The loss combines reconstruction fidelity with an \(\ell_{1}\) sparsity penalty,
\begin{equation}
\mathcal{L}_{\mathrm{SAE}}
      =\bigl\|\widehat{\mathbf{h}}-\mathbf{h}\bigr\|_{2}^{2}
      \;+\;\lambda\,\bigl\|\mathbf{z}\bigr\|_{1},
\end{equation}
where \(\lambda\) balances reconstruction accuracy against population sparsity. After convergence, the SAE parameters are frozen, and the normalized decoder rows \(\{\mathbf{w}_{k}\}_{k=1}^{d_{\text{hid}}}\) constitute a sparse feature dictionary.  Activating a single latent coordinate \(z_{k}\) moves an embedding exclusively along \(\mathbf{w}_{k}\), enabling controlled manipulations, causal ablations, and correlation studies with physical flow quantities.  Subsequent analyses can be done by using these properties to relate latent directions to interpretable CFD phenomena without altering the surrogate’s predictive pipeline.

\section{Experimental Setting}
\subsection{Dataset}
In this work, we utilize the CylinderFlow dataset \cite{pfaff2020learning}, which includes simulations of transient incompressible flow around a cylinder, with varying diameters and locations, on a fixed 2D Eulerian mesh. In all fluid domains, the node type distinguishes fluid nodes, wall nodes, and inflow/outflow boundary nodes. The inlet boundary conditions are given by a prescribed parabolic profile. The dataset contains 1000 training simulations, 100 validation simulations, and 100 test simulations, with each simulation spanning 600 time steps. 
\subsection{Training Details}
The MGN surrogate is trained using the same hyperparameter configuration as described in the original MeshGraphNet paper~\cite{pfaff2020learning}. Specifically, the network performs nine message-passing iterations with a latent dimension of $128$ for both node and edge features. Each update module consists of a residual MLP with two hidden layers and employs layer normalization. 

The SAE is trained post-hoc on the frozen node embeddings obtained from the trained MGN model. We use an expansion factor of $\kappa = 8$, resulting in a hidden layer width of $d_{\mathrm{hid}} = 8d_{\mathrm{in}}$ for input dimension $d_{\mathrm{in}} = 128$. The sparsity coefficient is set to $\lambda = 3 \times 10^{-4}$, and training is performed with a mini-batch size of $128$ using the Adam optimizer. Training proceeds until the reconstruction loss on a held-out validation set stops decreasing.

\subsection{Selecting Salient Latent Dimensions}
\label{sec:salient_selection}
Although the sparse autoencoder expands every frozen MGN embedding 
$\mathbf{h}_{i,t}^{L}\!\in\!\mathbb{R}^{d_{\text{in}}}$ 
to a $d_{\text{hid}}(=\kappa d_{\text{in}})$‑dimensional code
$\mathbf{z}_{i,t}\!=\!(z_{i,t}^{(1)},\dots ,z_{i,t}^{(d_{\text{hid}})})$,  
not all dimensions are activated and have meaningful information stored.
For analysis, we focus on retaining the Top‑$K$ salient coordinates.
\begin{equation}
\mathcal K=\bigl\{\,k_{1},\dots ,k_{K}\bigr\}\subset\{1,\dots ,d_{\text{hid}}\},
\qquad K\ll d_{\text{hid}}.
\end{equation}
Focusing on \(\mathcal{K}\) offers three advantages:  
\textit{(i)} it enables low‑dimensional visualisations that can be linked
directly to physical flow structures;  
\textit{(ii)} it isolates the latent directions that drive the dominant
dynamical variations; and  
\textit{(iii)} it reduces the computational load for any downstream
post‑hoc model that consumes SAE codes.

To construct \(\mathcal{K}\), we assign every dimension
\(d\in\{1,\dots,d_{\mathrm{hid}}\}\) a scalar relevance score \(s_{d}\) and
retain the \(K\) largest.  We consider three complementary scoring criteria whose closed‑form
definitions are summarized in Table~\ref{tab:ranking_criteria}.
\begin{table}[t]
\centering
\footnotesize        % smaller font
\setlength{\tabcolsep}{4pt}  % tighter column spacing
\caption{Closed‑form scores used to rank each latent dimension \(d\).  
\(z_{i,d}\) is the activation of dimension \(d\) in sample \(i\);
\(M\) is the number of samples;  
\(p_{b,d}\) is the empirical probability of \(z_{*,d}\) falling in histogram bin \(b\) (with \(B\) bins). Higher scores correspond to dimensions that are more informative under the stated interpretation.}
\label{tab:ranking_criteria}
\vskip -0.15in
\begin{tabular}{@{}p{0.15\linewidth} l p{0.36\linewidth}@{}}
\toprule
\textbf{Criterion} &
\multicolumn{1}{c}{\textbf{Score \(s_d\)}} &
\textbf{Interpretation} \\ \midrule
\textbf{Variance} &
\(\displaystyle s_d^{\mathrm{var}}=\frac{1}{M}\sum_{i=1}^{M}\bigl(z_{i,d}-\bar z_d\bigr)^{2}\) &
Spread around the mean \\
\textbf{Mean Absolute} &
\(\displaystyle s_d^{\mathrm{abs}}=\frac{1}{M}\sum_{i=1}^{M}\lvert z_{i,d}\rvert\) &
Average activation magnitude \\
\textbf{Entropy} &
\(\displaystyle s_d^{\mathrm{ent}}=-\sum_{b=1}^{B} p_{b,d}\log p_{b,d}\) &
Distributional diversity \\ \bottomrule
\end{tabular}
\end{table}
% \begin{itemize}
% \item \textbf{Variance}
% \begin{equation}
% s_{d}^{\text{var}}
%       \;=\;
%       \frac{1}{M}\sum_{(i,t)}
%       \Bigl(z_{i,t}^{(d)}-\overline{z}^{(d)}\Bigr)^{2},
% \label{eq:variance}
% \end{equation}

% \item \textbf{Mean absolute activation}
% \begin{equation}
% s_{d}^{\text{abs}}
%       \;=\;
%       \frac{1}{M}\sum_{(i,t)}
%       \bigl|\,z_{i,t}^{(d)}\bigr|,
% \label{eq:meanabs}
% \end{equation}

% \item \textbf{Shannon entropy}
% \begin{equation}
% s_{d}^{\text{ent}}
%       \;=\;
%       -\sum_{b=1}^{B}p_{b,d}\,\log p_{b,d},
% \qquad
% p_{b,d}=\frac{\#\bigl\{z_{i,t}^{(d)}\!\in\!\text{bin }b\bigr\}}{M},
% \label{eq:entropy}
% \end{equation}
% \end{itemize}
% where $M$ is the number of $(i,t)$ samples considered, 
% $\overline{z}^{(d)}$ is the sample mean,
% and the histogram in \eqref{eq:entropy} uses $B$ equally spaced bins.
% Higher scores indicate greater information content.

\subsection{Global vs. Time-Local Saliency in Latent Space}
\label{sec:two_regimes}
Because CFD rollouts evolve in time, a single ranking criterion can blur the
distinction between \textit{persistent} and \textit{transient} latent features.
We therefore extract salient dimensions at two complementary temporal scales.
\\
\textbf{Global\;($\mathcal K^{\mathrm G}$)}: 
      Pool all snapshots, compute a single score $s_{d}^{\mathrm G}=g\bigl(\{z_{i,t}^{(d)}\}_{i,t}\bigr)$
      with $g$ as the score functions in Table~\ref{tab:ranking_criteria},
      and keep the $K$ best:
      $
      \mathcal K^{\mathrm G}
        \;=\;
        \operatorname{top}_{K}\bigl(\{s_{d}^{\mathrm G}\}_{d=1}^{d_{\text{hid}}}\bigr).
      $\\
\textbf{Time-Local\;($\mathcal K^{\mathrm T}_{t}$)}: 
      Score each snapshot separately,
      $s^{(t)}_{d}=g\bigl(\{z_{i,t}^{(d)}\}_{i}\bigr)$,
      and obtain a {time‑indexed} set
      $
      \mathcal K^{\mathrm T}_{t}
        \;=\;
        \operatorname{top}_{K}\bigl(\{s_{d}^{(t)}\}_{d=1}^{d_{\text{hid}}}\bigr) \text{ with } t=1,\dots ,T.
      $

Global selection uncovers dimensions that are consistently important
throughout the rollout;
the time-local view reveals transient factors that spike during specific
flow events (e.g., vortex shedding).  
%Section~\ref{sec:compare-regimes} quantifies and visualizes the relationship
%between $\mathcal K^{\mathrm G}$ and $\{\mathcal K^{\mathrm T}_{t}\}$.

\subsection{Spatial Localization of Salient Latent Dimensions}

After identifying the Top‑\(K\) latent dimensions, we determine where on the mesh
those dimensions are most active for better illustration. Let \(\mathbf{Z}\in\mathbb{R}^{T\times N\times d_{\mathrm{hid}}}\) be the
tensor of SAE activations, where \(T\) is the number of snapshots,
\(N\) the number of mesh nodes, and \(d_{\mathrm{hid}}\) the size of the
sparse dictionary.  For each time step \(t\), we compute a node‑level score
$a_{i,t} \;=\; \sum_{d\in\mathcal K_{t}} z^{(d)}_{i,t} \text{ with }i = 1,\dots,N,
$
where \(\mathcal K_{t}\) is either the \textit{global} set
\(\mathcal K^{\mathrm G}\) (constant in \(t\)) or the
\textit{time-local} set \(\mathcal K^{\mathrm T}_{t}\)
defined in Section~\ref{sec:two_regimes}.
The score \(a_{i,t}\) is the cumulative activation a node receives from all
selected latent dimensions at snapshot \(t\). This simple aggregation avoids cancellations across dimensions and highlights mesh regions that concentrate the bulk of
the latent activity, thereby linking abstract SAE features to concrete,
spatial flow structures.

\begin{figure}[htbp]
    \centering
    \includegraphics[width=0.45\textwidth]{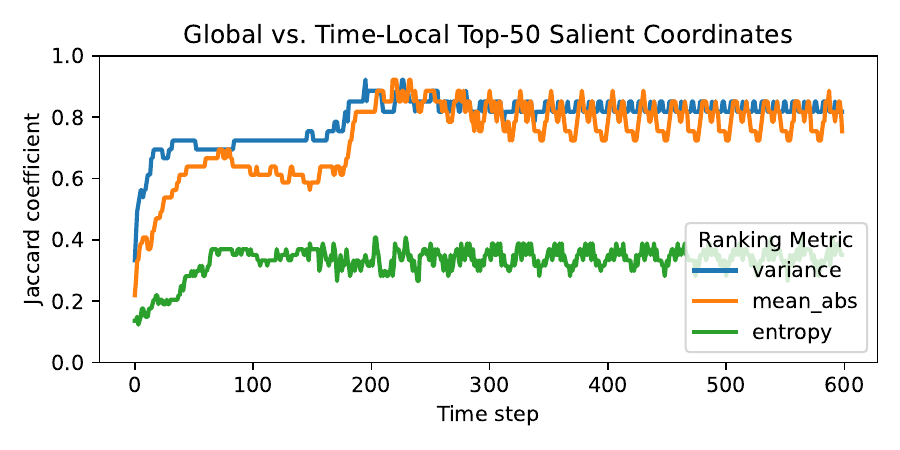}
    \vskip -0.15in
    \caption{Jaccard similarity between global and time-local Top‑\(50\) latent dimensions for a selected testing case, evaluated with three  ranking metrics. The periodic oscillations in the \textit{variance} and \textit{mean absolute} curves exhibit clear shedding‑period oscillations, indicating that most globally salient features remain important at each snapshot while alternating with transient modes. The \textit{entropy} metric also follows the shedding rhythm, but at a much lower mean level, showing that it continually turns over a larger share of dimensions and therefore captures more short‑lived, fine‑scale structures.}
    \label{fig:jaccard-rank}
\end{figure}

\section{Result and Analysis}

\subsection{Relation Between Global and Time-Local Top-\(K\) Salient Coordinates}

The two extraction regimes introduced in
Section~\ref{sec:two_regimes} serve complementary purposes:
\textit{global} ranking \(\mathcal K^{\mathrm G}\) finds dimensions that
matter throughout the rollout, whereas the \textit{time‑local} sets
\(\{\mathcal K^{\mathrm T}_{t}\}\) expose short‑lived,
event‑specific features.
A ranking metric (as introduced in Section \ref{sec:salient_selection}) is most useful for \textit{post‑hoc interpretation} when the
globally selected coordinates also remain salient at individual snapshots.
To quantify this alignment, we compute, for each time step,
the Jaccard similarity between the two Top‑\(K\) sets,
\begin{equation}
J(t)=
\frac{\bigl|\mathcal K^{\mathrm G}\cap \mathcal K_{t}^{\mathrm T}\bigr|}
     {\bigl|\mathcal K^{\mathrm G}\cup \mathcal K_{t}^{\mathrm T}\bigr|},
\qquad
J(t)\in[0,1],
\label{eq:jaccard}
\end{equation}
which equals \(1\) when the selections coincide and \(0\) when they are
disjoint.  Plotting \(J(t)\) (e.g. Figure~\ref{fig:jaccard-rank}) reveals how
stable each metric’s Top‑\(K\)  features are over time.

\subsection{Mesh‑Space Visualization of Aggregated Latent Activity}
\label{sec:mesh_agg}
\begin{figure}[htbp]
    \centering
    \includegraphics[width=0.48\textwidth]{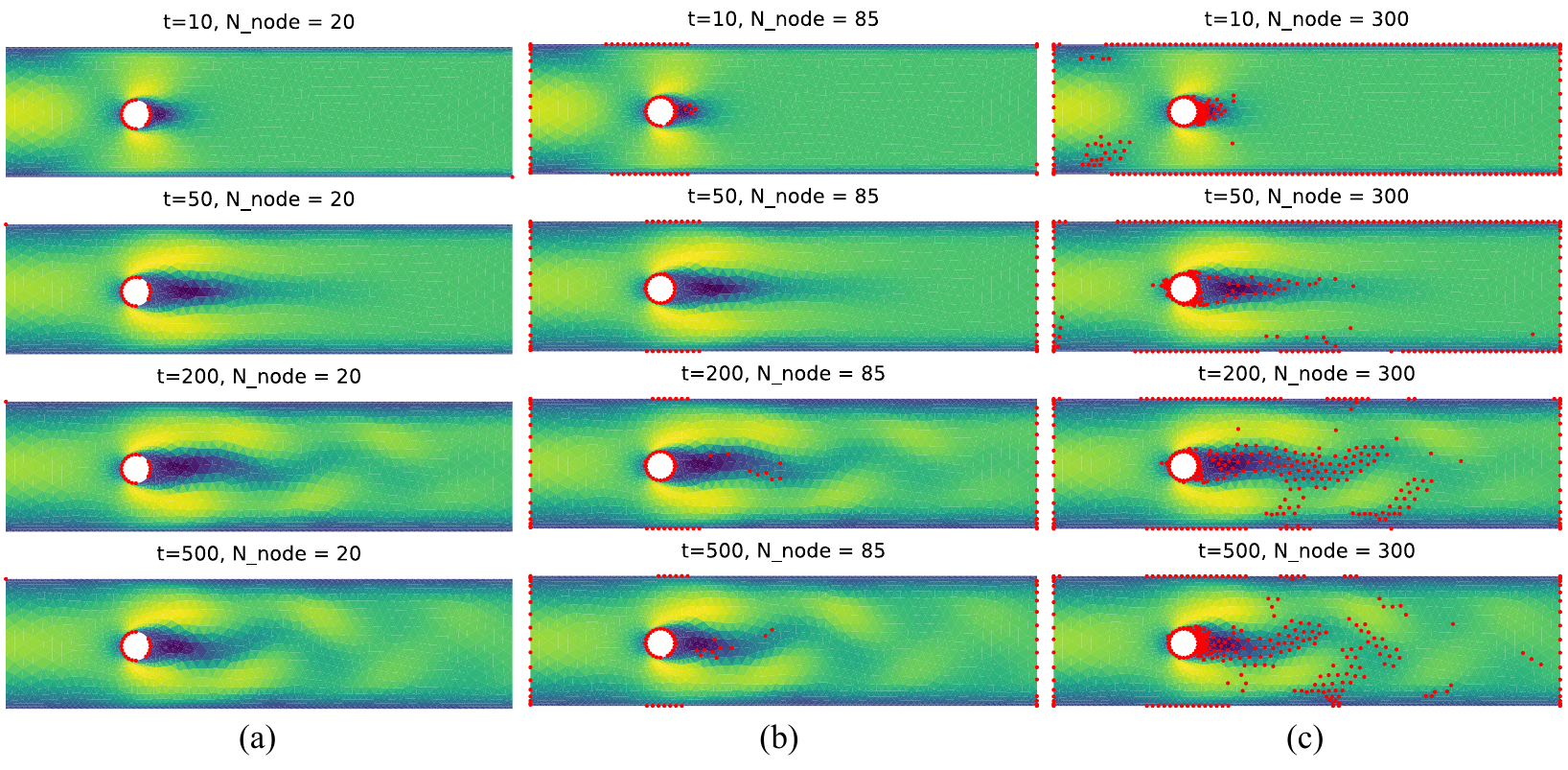}
    \vskip -0.15in
    \caption{Mesh–space footprint of aggregated salient latent
    dimensions selected via the variance metric. Red markers denote the top \(\eta\) nodes with the highest cumulative activation at four
    snapshots.  }
    \label{fig:mesh_variance_global}
\end{figure}

For each snapshot, we aggregate the activations of the globally selected
latent dimensions (Section~\ref{sec:two_regimes}) and rank the nodes
by their cumulative score
$
    \mathcal N_{t}
    \;=\;
    \operatorname{top}_{\eta}
    \bigl(\{a_{i,t}\}_{i=1}^{N}\bigr),
$
where \(\eta\) is a user-defined sampling budget.
Figure~\ref{fig:mesh_variance_global} overlays the \(\eta\) most active
nodes (red) on the CFD mesh for three budgets and four representative
time steps.

The selected nodes reveal a clear, physics‑consistent hierarchy. The selection first concentrates on the cylinder surface (\(\eta=20\)), then spreads to inlet, outlet, and boundary walls (\(\eta=85\)), and finally fills the wake (\(\eta=300\)), mirroring the physical hierarchy of importance in cylinder flow. This systematic outward expansion shows that the sparse autoencoder captures coherent, hierarchically ordered flow features, underscoring the interpretability of the learned MGN surrogate.
    
% With the strictest budget (\(\eta=20\)), the hotspots cluster on the cylinder
% surface, pinpointing regions of boundary‑layer separation and high pressure
% variance—precisely where the flow first departs from symmetry.
% Increasing the budget to \(\eta=85\) adds wall and inlet nodes that regulate
% the global pressure field, while the largest budget (\(\eta=300\)) extends
% into the downstream wake, marking the alternating vortex cores of the
% Kármán vortex street.

\vspace{-0.2in}

\subsection{Mesh‑Space Visualization of Individual Latent Activities}

Unlike the aggregated hotspots in Section~\ref{sec:mesh_agg}, which show where the surrogate concentrates its collective attention, visualizing one latent dimension at a time exposes the unique spatial footprint of each latent dimension, confirming that different dimension specialize in distinct flow structures and thereby providing a more granular, physically interpretable view of the model's internal representation via the SAE.

\begin{figure}[htbp]
    \centering
    \includegraphics[width=0.48\textwidth]{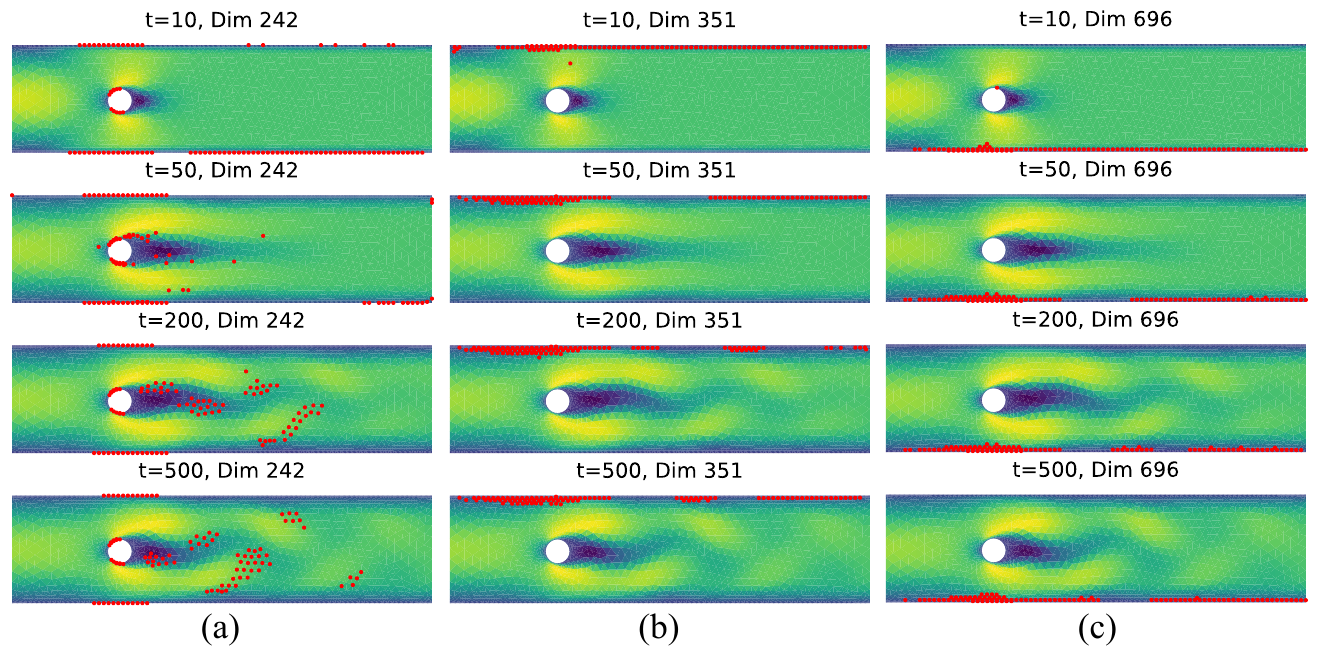}
    \vskip -0.15in
    \caption{Spatial footprint of three individual salient latent dimensions selected via the mean‑absolute metric. Red markers show
    the \(\eta\) most active nodes (\(\eta=100\)) at four time steps.  Each dimension localizes to a distinct flow feature.}
    \label{fig:mesh_mean_abs_individual_dim}
\end{figure}

For each selected dimension \(d\in\mathcal K\), we rank nodes by the single‑dimension activation \(a^{(d)}_{i,t}=z^{(d)}_{i,t}\) and highlight the top
\(\eta\) nodes on the mesh. Figure~\ref{fig:mesh_mean_abs_individual_dim} highlights three representative dimensions from the mean‑absolute Top‑10 representative salient dimensions and visualizes across four snapshots. The clear spatial disjointness of these dimensions confirms that the SAE dictionary disentangles disparate physical phenomena, underscoring the surrogate’s interpretability and enabling feature‑specific diagnostics or control.

\subsection{Quantitative Evaluation on Vortex Alignment}
To quantitatively compare post‑hoc interpretability, we treat the top 10\% vorticity nodes (calculated using the method described in \cite{holmen2012methods}) at each snapshot as ground truth and evaluate how frequently each method’s top‑$\eta$ mesh node selections fall within this mask. We report precision, recall, F\(_1\), and Jaccard, averaged over time, for three SAE
variants and three baseline methods: \textit{(a)} \textbf{\textit{Embedding-norm}}: ranks nodes by \(\|\mathbf h^{L}_{i,t}\|_{2}\); \textit{(b)} \textbf{\textit{PCA}}: projects embeddings onto the first $m = \lfloor K / \kappa \rfloor$ principal components and ranks by projection norm;  \textbf{(c)} \textbf{\textit{Random}}: uniform node sampling.

Table \ref{tab:vortex_scores} shows that variance and mean-absolute scoring yield the highest scores, whereas the entropy metric ranks lowest among SAE variants—yet still outperforms all non-SAE baselines. This agrees with Fig.\ref{fig:jaccard-rank}, where the entropy curve exhibits lower values and larger oscillations than its counterparts, suggesting that this metric emphasizes transient features and consequently captures a smaller portion of the coherent vortex region. Overall, sparse autoencoding provides a more precise and physically grounded node-saliency estimate than magnitude- or subspace-based heuristics.

%SAE\(_{\text{variance}}\) yields the highest scores, with the other two SAE criteria close behind; all three outperform every non‑SAE baseline.  Although absolute numbers are low—each method must choose only \(\eta\) nodes out of thousands while the mask covers just 5\% of the mesh—the consistent gap confirms that sparse‑autoencoder saliency pinpoints physically meaningful, high‑vorticity regions better than magnitude‑based or linear subspace heuristics, making it the most reliable option for post‑hoc interpretability for our problem setting.

\begin{table}[t]
\centering\footnotesize
\caption{Quantitative comparison of node-level interpretability across methods. Scores reflect how well each method’s top-\(\eta\) nodes across top-$K$ latent dimensions align with high-vorticity regions, averaged over time. SAE variants consistently outperform baselines.}
\vskip -0.15in
\label{tab:vortex_scores}
\begin{tabular}{lcccc}
\toprule
Method              & Precision $\uparrow$ & Recall $\uparrow$ & F1 $\uparrow$    & Jaccard $\uparrow$ \\
\midrule
SAE (variance)      & {0.70}      & {0.52}   & {0.60}  & {0.43}    \\
SAE (mean abs)      & {0.70}      & {0.52}   & {0.60}  & {0.43}    \\
SAE (entropy)       & 0.64      & 0.48   & 0.55  & 0.38    \\
\midrule
Embedding-norm      & 0.64      & 0.48   & 0.55  & 0.38    \\
PCA                 & 0.57      & 0.43   & 0.49  & 0.32    \\
Random              & 0.10      & 0.08   & 0.09  & 0.05    \\
\bottomrule
%\vspace{-0.3in}
\end{tabular}
\end{table}

% As shown in Table \ref{tab:vortex_scores}, the \emph{SAE\(_\text{variance}\)} baseline attains the highest alignment with vortex nodes, followed by \emph{SAE\(_\text{mean\_abs}\)} and \emph{SAE\(_\text{entropy}\)}, all of which outperform the non-SAE baselines. Although the absolute values are modest due to the constrained node budget, the consistent lead of SAE demonstrates that sparse autoencoder selection more effectively identifies physically meaningful, high-vorticity regions than simpler heuristics. And thus better for post-hoc assessment.

\section{Conclusion}

We introduce the first post‑hoc interpretability pipeline for graph‑based CFD surrogates by training a sparse autoencoder on learned MGN embeddings. By ranking latent directions, examining their temporal stability, and projecting individual dimensions back to mesh space, we uncovered a compact, physically coherent vocabulary of flow features—boundary‑layer separation, wall effects, and wake vortices. Quantitative evaluation against high‑vorticity ground truth showed that SAE‑based saliency consistently outperforms embedding‑norm, PCA, and random baselines, confirming its ability to highlight dynamically relevant regions. These results demonstrate that sparse autoencoding is a practical and effective tool for explaining graph surrogates in fluid dynamics, offering a model-agnostic pathway to enhance explainability and trustworthiness in CFD applications.

%\section*{Acknowledgments}

\section*{Acknowledgments}
This work was performed under the auspices of the U.S. Department of Energy by Lawrence Livermore National Laboratory under Contract DE-AC52-07NA27344. The work was partially supported by the LLNL-LDRD (23-ERD-029) and DOE ECRP (SCW1885).

\bibliographystyle{named}
\bibliography{ijcai25}

\end{document}